\documentclass[aps,prl,superscriptaddress,amsmath,groupedaddress,twocolumn]{revtex4}
\usepackage{color,hangcaption,hhline,psfrag,rotating,amssymb}
\usepackage[hang,nooneline]{subfigure}
\usepackage{dcolumn}

\begin{document}
\title{Lattice QCD 2007}

\author{C. T. H. Davies}
\email[]{c.davies@physics.gla.ac.uk}
\affiliation{Department of Physics and Astronomy, University of Glasgow, Glasgow, G12 8QQ, UK}
%\thanks{}
%\affiliation{}

%\collaboration{HPQCD and UKQCD}
\noaffiliation

\date{\today}

\begin{abstract}
I review the current status of lattice QCD calculations 
and the progress that we have made in consolidating the 
`lattice QCD revolution' of four years ago. 
Significant results from formalisms other than the improved 
staggered formalism, which has been leading the revolution, are now
appearing. Comparison between formalisms gives additional 
confidence in the results.   

More 
precision tests against experiment have been made and 
predictions that are needed by the experimental programme 
have been improved. There has been particular progress 
this year in charm and strange physics. 
\end{abstract}

%\pacs{}
%\keywords{}

\maketitle

\section{Introduction}

QCD is a key component of the Standard Model of particle physics. On 
the one hand, it gives 
us a rich spectrum of bound states of quarks and gluons whose properties 
are predictable from QCD if we can solve the theory. On the other hand, the 
confinement of quarks complicates the determination of 
the properties of quarks from experiment because only hadrons can be 
studied directly. 
Lattice QCD enables QCD effects 
to be calculated `from first principles' in the hadronic regime where 
the theory is strongly-coupled and nonlinear. Accurate results can 
provide stringent tests of QCD when compared to experiment as well
as providing input to our understanding of the Standard Model in the quark sector. 

In current lattice QCD we can calculate the masses of `gold-plated' hadrons (those 
with small width well below Zweig-allowed decay thresholds) and simple 
decay matrix elements that include at most one gold-plated hadron in the final 
state. There are both statistical and systematic errors from lattice calculations, 
however, and it is important to understand the sources of these so that 
an optimal strategy can be developed to minimise the total error from the 
calculation.

One area in which lattice QCD is making an important contribution is that 
of flavor physics and the determination of elements of the 
Cabibbo-Kobayashi-Maskawa matrix. This is linked to the  
worldwide experimental programme determining 
weak decay and mixing rates for bottom and strange hadrons with a view 
to fixing all 3 sides of the unitarity triangle and so testing the 
internal consistency of the Standard Model's description of CP violation. 
The experimental programme will achieve errors of a few percent on the 
decay rates (and has already done this on mixing rates) 
and needs theoretical input for the Standard Model prediction 
to extract the appropriate CKM element. The final error for the CKM 
element, and corresponding side of the unitarity triangle, is 
currently limited 
by the error from lattice QCD and this must be reduced to be much 
closer to the experimental error, if that accuracy is not to be wasted. 

The good news is that it now looks possible to achieve this as a result 
of key advances over the past decade in understanding how to discretise 
QCD accurately onto a space-time lattice. Lattice QCD calculations are 
numerically extremely expensive, but the savings that have resulted from 
improved discretisation have at last brought realistic calculations within 
the power of current day supercomputers~\cite{highp}. I will outline the advances that 
have made the recent calculations possible and their implications for future 
work, concentrating on calculating relevant to the flavor physics and CKM programme. 
The Proceedings of this year's lattice conference should be consulted for a 
more general view~\cite{lat07}. 

\section{Lattice calculations}

Lattice QCD proceeds by the numerical evaluation of the Feynman path integral~\cite{book}. 
For this integral to be finite and well-behaved we work with a finite 
volume of 4-dimensional space-time in which time is rotated to (imaginary) Euclidean time. 
The space-time within our volume is split up into a lattice of points with 
lattice spacing $a$ and 
there are then a finite number of quark and gluon quantum fields residing 
on the sites (quarks) and links (gluons) of the lattice. 
We then have to integrate over all possible values of these fields, weighted 
exponentially by (minus) the action, $S$ (integral of the Lagrangian), 
of QCD. In practise this means 
using a random process to 
generate sets of possible gluon fields, one for every link of the lattice, 
called configurations. This is the `data generation' phase of 
a lattice QCD calculation. If we generate these configurations with probability 
$e^{-S}$ then we are preferentially choosing configurations that contribute 
most to the path integral and we can evaluate it efficiently. We call these 
configurations `typical snapshots of the QCD vacuum'. A set of configurations
for a particular set of QCD parameters is called 
an ensemble. An ensemble for a good calculation will generally have several 
hundred configurations in it. The evaluation, 
or analysis, stage of a lattice QCD calculation consists of `measuring' 
various functions of the gluon fields that correspond to a particular 
observable, such as a correlation function from which a hadron mass 
can be determined. The function of the gluon fields is evaluated on 
each configuration of the ensemble and the mean value and its statistical 
error determined. The statistical error will depend on the number of 
configurations in the ensemble, i.e. how well the path integral is 
approximated by this procedure. 
The statistical error for a `measurement' varies as the inverse square root of the number 
of configurations in the ensemble and to reduce this error to 1\% is perfectly feasible 
in current calculations for quantities such as flavour non-singlet ground-state 
hadron masses. Reducing systematic errors to this level is much harder, and it has 
taken many years of effort to understand how to do this.  

The key source of systematic error
is that coming from the 
discretisation of the Lagrangian of QCD onto a space-time lattice. 
Discretisation errors invariably arise when equations from continuous 
space-time are discretised for 
numerical solution. Typically physical 
results will depend on the unphysical step-size or spacing chosen for 
the discretisation. The physical result for continuous space-time 
will be obtained either by extrapolating in the step-size to zero, or 
reducing these errors to a known and acceptable level. 
The errors arise, for example, from the approximation of derivatives 
by finite differences, and for classical equations they are readily 
reduced by using higher-order 
differencing schemes which correct for the errors being made in 
low-order schemes by adding additional terms to cancel them. 
This has the effect of raising the power of the step-size at which 
errors first appear. This therefore reduces them, at fixed small step-size, 
or allows you to achieve you the same errors with a cheaper 
calculation with larger step-size. 

Exactly the same considerations apply in lattice QCD. 
The discretisation step-size is the separation between 
points in the lattice and is known as the lattice spacing, $a$. 
The number of lattice points required in a fixed physical 
volume grows as $a^{-4}$ as the lattice spacing is reduced.
However, the cost of a calculation grows much more rapidly than 
this, approximately like $a^{-7}$, because the efficiency 
with which statistically independent configurations can be generated 
also falls as a function of $a$. 
This makes it impossible to work at very small 
values of $a$ and hence understanding the systematic discretisation errors 
associated with the lattice spacing has been critical in enabling 
us to obtain results at values of $a$ at which we can afford 
to do the calculation. 

Discretisation of the gluon piece of the QCD action, including 
the effect of `improvement' to reduce its discretisation errors 
is relatively straightforward. It is the quark piece of the QCD action 
that causes the most headaches and controversy. 
Since quark fields anticommute, as they are fermions, they cannot be included directly in our 
numerical simulations with commuting arithmetic, but must be integrated 
out of the Feynman path integral. The path integral then becomes an integral 
over gluon fields {\it only}, with the effect of quarks being implicit 
as functions of the gluon field. If we write the quark piece of the 
QCD Lagrangian as 
\begin{equation}
L_{q, QCD} = \overline{\psi}(\gamma \cdot D + m) \psi = \overline{\psi} M \psi
\end{equation}
then we call $M$ the 'fermion matrix'. $\gamma$ are the Dirac $\gamma$ matrices, 
$D$ is the covariant derivative that includes coupling to the gluon field 
and $m$ is the quark mass. The quark field, $\psi$, is a 4-spin, 3-color 
vector on every site of the lattice so $M$ is a $12V \times 12V$ matrix where 
$V$ is the number of sites on the lattice. 
The result of integrating quarks out of the Feynman path integral means that 
their effects appear in two different ways. For valence quarks 
we must calculate the `quark propagator', $M^{-1}$, and combine quark 
propagators to make hadron correlation functions. For sea quarks, we need 
to include ${\rm det}(M)$ in the generation of gluon field configurations 
to make `typical snapshots of the vacuum' that include the effect of sea quark-antiquark
pairs being produced by energy fluctuations in the vacuum. 

The calculation of rows of $M^{-1}$ is numerically costly because $M$ is such a large 
matrix. It is also ill-conditioned as $m \rightarrow 0$ and this is unfortunate 
since we have quarks in nature, the $u$ and $d$ quarks, that are very light. 
The inclusion of ${\rm det}(M)$ in the generation of gluon field configurations 
is numerically even more costly since it involves many calculations of 
rows of $M^{-1}$. Here the fact that the $u$ and $d$ quarks are so light 
(and $s$ to a lesser extent) is 
even more of a problem, but at the same time it is exactly the reason 
that they are the most physically important as sea quarks since they 
can readily be generated by a vacuum energy fluctuation.  

The quarks that we need to be concerned with are only the 5 lightest, 
because the $t$ quark does not form hadronic bound states. $c$ and $b$ 
quarks are themselves too heavy to have any significant effect as 
sea quarks, and appear in existing calculations only as valence quarks. 
$u$, $d$, and $s$ sea quarks are physically important, however, 
as stated above. 
Their inclusion is necessary, for 
example, for the strong coupling constant to run correctly. Without this, 
quantities which are sensitive to different energy scales will not 
agree. Early lattice 
calculations did not have the computer power to include sea quarks 
and so they were dropped in what was known as the `Quenched Approximation'. 
This approximation has systematic errors at the 10\% level which 
can be moved around between different quantities depending on how 
the calculation is done, because it is not an internally consistent 
approximation to QCD. 
Subsequently calculations including sea quarks were done, but they 
included only two flavours of sea quarks (i.e. $u$ and $d$) but with 
masses that were 10-20 times too big (i.e. around the mass of the strange 
quark).  
The numerical cost of including light $u$ and $d$ quarks means that 
it is very important to find a fast and accurate discretisation of the 
quark Lagrangian so that affordable calculations can be done at moderate values 
of $a$ (around 0.1fm). In practice calculations are still done at 
multiple values of the $u$ and $d$ quark masses that are too heavy and 
then extrapolations are made to the physical point. These extrapolations 
can be guided by chiral perturbation theory~\cite{book}, which gives an expansion in powers
of the $u/d$ quark masses, provided that we are close enough to $m_{u/d}=0$.  

Controversy enters at this point since there are many different formulations 
for quarks in lattice QCD, depending on the approach taken to solving 
the infamous `doubling problem'. When the Dirac Lagrangian above is 
discretised onto the lattice it describes $2^d$ continuum quarks in $d$ 
dimensions rather than 1. This gives 16 quarks in 4 dimensions. We call 
the 15 additional quarks `doublers' or additional `tastes' of quark. Different 
formulations take a different attitude to this problem with consequent effects 
on speed and discretisation errors. Maintaining as much as possible of the chiral symmetry 
of QCD, under which left- and right-handed projections of the quark 
field can be separately rotated in 
the absence of a quark mass, is also important. The pion mass is guaranteed 
to vanish at zero quark mass, for example, because it is the Goldstone boson of 
spontaneously broken chiral symmetry. If this can be maintained on the 
lattice it is a big advantage in terms of easily being able to find your 
way, from a given lattice quark mass, to the physical point for $u$ and 
$d$ quarks, close to zero quark mass. In some lattice quark formalisms 
this property has to be given up and then the value of the quark mass (which can be negative) 
corresponding to zero pion mass has to be searched for, which is an added 
complication. On the other hand maintaining the best possible version of continuum 
chiral symmetry on the lattice is numerically extremely expensive and not 
a particular advantage for a lot of calculations. 

Table~\ref{tab:formalism} gives a 
brief outline of the existing formalisms for which significant numbers of gluon 
field configurations including 
the effect of sea quarks have been made, and a significant amount of physics analysis 
done. It includes the names of the collaborations 
chiefly using those formalisms. 
It is also possible to use a different valence quark formalism 
from that used for the sea quarks included in the gluon field configurations that 
you are working on. For example, several collaborations have used domain wall 
valence quarks on the MILC collaboration configurations with improved staggered 
sea quarks. The table briefly describes the good and bad points of each formalism 
with respect to speed and chiral symmetry. There are other technical issues associated 
with each formalism that the collaborations involved have spent considerable time 
understanding. In most cases they amount to restrictions on how the continuum ($a \rightarrow 0$)
and chiral ($m_q \rightarrow 0$) limits are approached, and they limit how small 
a value can be taken for $m_q$ at finite $a$. For example, for improved staggered quarks 
no attempt is made to solve the doubling problem and the doublers are taken 
care of by effectively 'dividing by 16'.
Away from the $a=0$ 
limit that this amounts to a discretisation error that can cause problems if the 
valence quark mass is taken too light. This has been demonstrated both 
numerically and in `staggered chiral perturbation theory', which takes account 
of the effect of the doublers in chiral perturbation theory. 
For a recent review of issues for staggered quarks see~\cite{kronfeld}. The discretisation 
errors are reduced further in a new highly improved staggered quark (HISQ) formalism 
introduced recently by the HPQCD collaboration~\cite{hisq}, that will be discussed further 
below. For domain 
wall quarks a small additive quark mass renormalisation appears which means
that care must again be taken with small valence quark masses.  For a recent review of 
issues for domain wall quarks see~\cite{boyle} and for the related overlap quarks 
see~\cite{matsufuru}. For twisted mass, see~\cite{urbach} and for clover~\cite{kuramashi}. 
In practice these issues are under control in existing calculations. Another
important problem, and one on which there has been a lot of recent work, 
is that of the lattice volume. 

The dependence on $m_{u/d}$ in chiral perturbation is obtained by integrating 
over the momenta of virtual pions. On the lattice the momenta are restricted 
to the discrete set available on a particular spatial volume. The lowest 
momentum (from discretising a derivative as a simple symmetric finite difference) 
in units of the lattice spacing is given by $pa = \sin(2\pi/L)$ where 
$L$ is the number of lattice points on a side of the lattice. Thus the smallest 
momentum (and therefore the infrared cut-off on the integral) is inversely 
proportional to the lattice size. At large quark masses the integral is 
cut-off by the quark mass but at small quark masses it becomes 
sensitive to the volume and this will distort the behaviour as a function of quark 
mass. This means that care must be taken at small 
$u/d$ quark masses to be working on a large enough spatial volume (that 
can be estimated using finite volume chiral perturbation theory~\cite{colangelo}), or 
to have multiple volumes and check the volume dependence explicitly (see, for 
recent examples~\cite{urbach, bernard}). Larger
spatial volumes are more expensive numerically, of course, since at fixed
lattice spacing they require more lattice points. 

\begin{table}
\begin{tabular}{l|l|l|l}
 & speed & chiral & collaborn \\
& & symmetry & \\
\hline
Improved & fast & OK & MILC/ \\
staggered &  & & HPQCD/ \\
(asqtad) &  & & FNAL \\
\hline
domain & slow & good & RBC/ \\
wall (DW) &  & & UKQCD \\
\hline
clover & fast & poor & PACS-CS \\
 &  & & QCDSF \\
 &  & & CERN-TOV \\
\hline
twisted & fast & OK & ETMC \\
mass &  & &  \\
\end{tabular}
\caption{ Different quark formalisms currently being used 
for generating gluon field configurations including the 
effect of sea quarks. Collaborations currently using these 
formalisms are given on the right.  
}
\label{tab:formalism}
\end{table}

\begin{figure}[h]
\rotatebox{270}{\includegraphics[height=10.0cm]{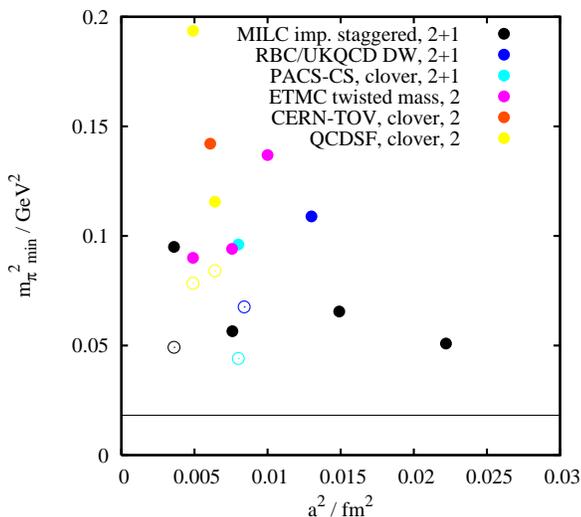}}
\caption{Configurations made by various collaborations using 
different quark formalisms and including the effects of 
$u/d$ sea quarks only ($n_f=2$) or $u,d$ and $s$ ($n_f=2+1$). 
The axes are chosen so that is clear how close to the `real world' 
(circled at $a=0$ and $m_{\pi}=0$) the configurations are. 
The $x$ axis gives $a^2$ since systematic errors for the 
actions shown here appear first at that order in $a$. 
The $y$ axis gives $m_{\pi}^2$ for $\pi$ mesons made 
of quarks with the same quark mass as those in the sea for 
the ensemble with lightest sea quark mass at that value of $a$. 
This is a measure of how far from the chiral limit the 
configurations are since $m_{\pi}^2$ is proportional 
to the light quark mass. Other considerations for the 
quality of an ensemble are the spatial volume of the 
lattices and how many independent configurations there 
are. Open symbols give parameters for configurations currently being generated. }
\label{fig:configstatus}
\end{figure}

Figure~\ref{fig:configstatus} shows the status of configurations 
that exist and are being generated as reported at the Lattice 2007 
meeting~\cite{lat07}. Improved staggered quarks have the best coverage of different 
values of the lattice spacing, but other formalisms are making good 
progress in generating configurations with light $u/d$ masses on relatively 
fine configurations. Not all of these include sea $s$ quarks at present
but all have plans to do so. 

From the very brief discussion above it will be clear that several different 
formalisms to handle quarks on the lattice are possible and different collaborations 
have strong views on the optimal approach. Of course, all the formalisms should 
give the same physical result in the end, and agreement between different 
formalisms is valuable confirmation of the validity of lattice QCD. 
Different formalisms have particular strengths and so it tends to be true that 
collaborations using different formalisms have a different focus for their physics 
programmes. 

\begin{figure}[h]
\includegraphics[height=3.0cm]{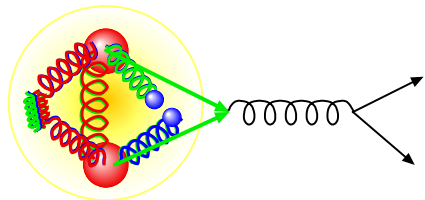}
\caption{ Annihilation of a charged pseudoscalar meson to a $W$ 
boson. The rate for this can be measured experimentally and depends 
on the appropriate CKM element coupling the annihilating valence 
quark and antiquark and on the `probability for the quark and 
antiquark to be in a position inside the meson to annihilate'. This 
probability is square of the matrix element of the axial vector 
current between the meson and vacuum, parameterised by $f_{\pi}$. 
It must be calculated 
in lattice QCD because it is sensitive to the long distance QCD 
effects that confine the quarks in the meson.  
}
\label{fig:2pt}
\end{figure}

There is overlap, however, in a lot of very basic calculations that act as 
a test case for the results. One example is the calculation of the 
pion decay constant $f_{\pi}$ that parameterizes the purely leptonic decay 
of a charged $\pi$ to leptons via a $W$ boson (see figure~\ref{fig:2pt}). 
The leptonic decay rate is proportional to the square of $f_{\pi}$ multiplied
by the square of $V_{ud}$ which is the appropriate CKM element by which 
the $u$ and $\overline{d}$ in the $\pi^+$ couple to the $W$. Given a very 
accurate value 
for $V_{ud}$ from superallowed $\beta$ decay it is possible to determine 
$f_{\pi}$ from the experimental leptonic decay rate and compare lattice QCD
results to this. The calculation of $f_{\pi}$ is relatively simple in 
lattice QCD for formalisms that have enough chiral symmetry to have a partially 
conserved axial current (so that no renormalisation is required). 
Then the calculation of $f_{\pi}$ is straightforwardly 
obtained from the same correlators that are used to obtain $m_{\pi}$.   
Calculations need to be done at several different values of the 
$u/d$ quark mass and the lattice spacing $a$.  For light enough 
$m_{u/d}$ a fit of the results to chiral perturbation theory can be 
done that allows an extrapolation to the physical value of $m_{u/d}$ 
(that gives the physical value of $m_{\pi}$). The different values of 
$a$ allow for a simultaneous extrapolation to the continuum 
limit.  At the physical point a comparison can be made with the 
value obtained from the experimental rate.  Using improved staggered quarks
or HISQ agreement with experiment is obtained with 1.5\% errors~\cite{hisqf}. This year 
new results from the twisted mass collaboration also gave 1\% accurate 
$f_{\pi}$ values (without fixing the lattice spacing) 
in the continuum and chiral limits with 2 flavors of 
sea quarks at a range of $m_{u/d}$ and with two values of the lattice spacing~\cite{twmf}. 

\begin{figure}[h]
\rotatebox{270}{\includegraphics[height=10.0cm]{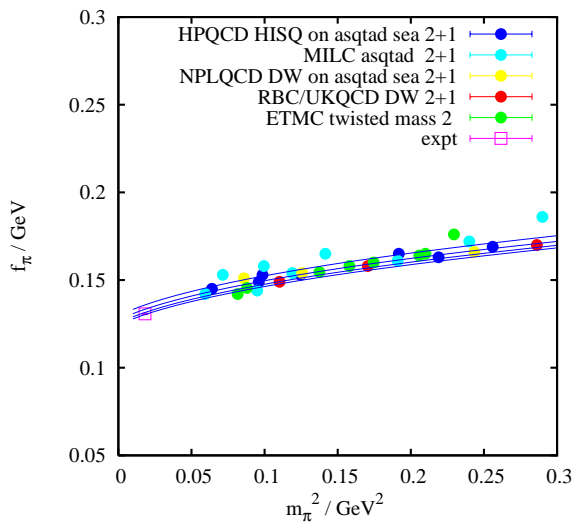}}
\caption{ A comparison of results for the pion decay constant 
obtained from different quark formalisms in lattice QCD 
including the effect of sea quarks. The results are plotted against
$m_{\pi}^2$ (which is proportional to $m_{u/d}$) and each calculation 
needs results at multiple values of $a$ and $m_{u/d}$ so that an 
extrapolation to the physical point where $m_{u/d}$ is small and 
$a=0$ can be made. Some formalisms have results only at one 
value of $a$ so far. The lines give the chiral extrapolation for 
the results from using the highly improved staggered quark formalism 
(HISQ) on the MILC ensembles using improved staggered quarks 
at each of 3 values of $a$ plus the $a=0$ line obtained~\cite{hisqf}. 
The results show very encouraging agreement between the different 
formalisms. The experimental point marked uses the leptonic decay 
rate of the $\pi$ meson and $V_{ud}$ from the particle data 
tables obtained from super-allowed $\beta$ decay. 
}
\label{fig:fpicomp}
\end{figure}

A comparison of different quark formalisms can also be made away from 
the chiral and continuum limits if discretisation errors are small. This 
comparison for formalisms with good chiral symmetry is shown in 
Figure~\ref{fig:fpicomp} and shows very encouraging agreement.
To make a more detailed comparison would require more work to remove 
effects of finite volume that different groups have handled in 
different ways. $f_{\pi}$ is sensitive to finite volume effects at
the lower end of the quark masses shown here and this has been 
explored by explicit calculations for the MILC~\cite{bernard}, ETMC~\cite{urbach} 
and RBC/UKQCD~\cite{juttner} results 
shown here, and compared to chiral perturbation theory expectations. 
The conclusion is that the effects are of the size you would expect  
- for example a 2\% shift downwards of $f_{\pi}$ on a 2.5fm lattice 
size (compared to the infinite volume real world) for $m_{\pi} \approx 250$ MeV. 
In Figure~\ref{fig:fpicomp} the MILC lattices at that $m_{\pi}$ are 2.8fm, 
the RBC/UKQCD lattices are 2.7fm (using their $24^3$ lattices rather than 
earlier $16^3$ ones) and the ETMC lattices are 2.2fm. 
The NPLQCD results use the MILC coarse lattices with domain wall 
valence quarks~\cite{nplqcd}.
Results with the clover formalism are also available but do not at present 
seem to agree as well, perhaps because of renormalisation issues~\cite{mcneile}. 
It is very exciting that we are now able to do a comparison between 
formalisms at this level of precision and more of this will become possible 
in future years. 

Handling heavy quarks on the lattice raises rather different issues from that 
for light quarks. If we use one of the standard light quark formalisms 
discretisation errors will be set by powers of $m_Qa$ where $m_Q$ is the heavy 
quark mass, rather than, as is typical for light quark quantities, powers 
of $\Lambda_{QCD}a$. For the $b$ quark $m_ba >> 1$ on typical lattices 
and so no amount of improvement can control the discretisation errors. 
However, we can make use of the fact that the $b$ quark is nonrelativistic 
inside its bound states and that $m_b$ is just an overall mass scale that 
does not affect the internal dynamics very much. The HPQCD collaboration 
has done a lot of work on bottomonium and $B$ physics using the nonrelativistic 
effective theory called NRQCD discretised on the lattice~\cite{nrqcd}. An alternative 
is to expand in powers of $1/m_Q$ away from the infinite quark mass (static) limit 
in the HQET approach to $B$ physics. The FNAL/MILC collaboration use a 
relativistic formalism (the clover formalism) making use of nonrelativistic 
understanding to remove key discretisation effects~\cite{fnalhq}. Again all the different 
formalisms should give the same physical results. For $c$ quarks the 
situation is less clear because on typical lattices $m_ca \approx 0.5$. 
The FNAL/MILC collaboration have made strong use of their clover `Fermilab' 
formalism for charmonium and $D$ physics. The HISQ formalism
uses a variant of the improved staggered formalism 
to reduce further the discretisation errors associated with multiple tastes. 
Other more standard discretisation are already removed to a very high level 
for this action and so it gives excellent results for $c$ physics that will 
be discussed below. It can also, of course, be used for $u$, $d$ and $s$ quarks, 
both as valence quarks (see below for results) and in the sea (in progress) 
as a further check that lattice 
discretisation errors are well understood. 

\section{Lattice Results 2007}

As discussed in the introduction lattice QCD has an important 
contribution to make to determining the elements of the CKM 
matrix. For each CKM element there is a gold-plated electroweak 
decay or mixing process whose rate, as for $\pi$ leptonic decay, will 
be (up to known kinematic factors) the product of that $V_{ab}^2$ and 
the square of a lattice QCD amplitude given as a decay constant, 
form factor or bag parameter that expresses the probability of 
the quarks confined inside the meson undergoing that process. 
The CKM matrix is given in Figure~\ref{fig:ckm} with the corresponding leptonic, 
semileptonic and mixing processes for each element.

\begin{figure}[h]
\[ \left( \begin{array}{ccc}
{\bf V_{ud}}  & {\bf V_{us}} & {\bf V_{ub}} \\
\pi \rightarrow l\nu & K \rightarrow l \nu & B \rightarrow \pi l \nu \\
 & K \rightarrow \pi l \nu &  \\
{\bf V_{cd}} & {\bf V_{cs}} & {\bf V_{cb}} \\
D \rightarrow l\nu & D_s \rightarrow l \nu & B \rightarrow D l \nu \\
D \rightarrow \pi l\nu & D \rightarrow K l \nu &  \\
{\bf V_{td}} & {\bf V_{ts}} & {\bf V_{tb}} \\
\langle  B_d | \overline{B}_d \rangle  & \langle B_s | \overline{B}_s \rangle  & \\
\end{array} \right) \]
\caption{ The CKM matrix with corresponding gold-plated processes that 
allow the value of each element to be determined by combining experiment 
with a lattice QCD calculation.}
\label{fig:ckm}
\end{figure}

\begin{figure}[h]
\rotatebox{270}{\includegraphics[height=10.0cm]{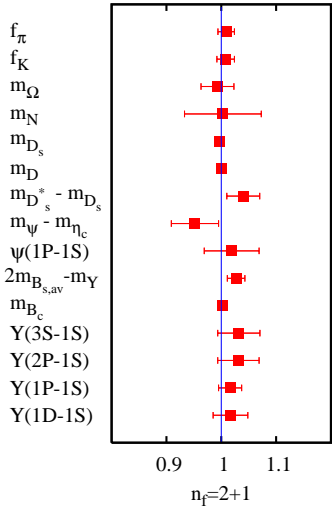}}
\caption{ The points show lattice QCD results divided by 
experiment for a range of quantities from light quark physics to 
bottomonium physics, compared to the right answer of 1.0. 
$f_{\pi}$ and $f_K$ are the $\pi$ and $K$ decay constants, 
described further in the text; $m_N$ is the nucleon mass. 
$\Upsilon(1P-1S)$ denotes, for example, the difference in 
mass between the lowest $\chi_b$ states (spin-averaged) 
and the $\Upsilon$. $\psi(1P-1S)$ is the same quantity 
for charmonium.  
The results come from analysis by the FNAL, HPQCD and MILC
collaborations on the MILC ensembles that include improved 
staggered sea $u$, $d$ and $s$ quarks~\cite{hisqf, hisq, toussaint, alan, allison, fnalpsi}. 
}
\label{fig:ratio}
\end{figure}

Of course, it is not sufficient to calculate only these processes in lattice 
QCD. It is important to have a number of cross-checks against other processes 
that are similar and well-known experimentally, for example electromagnetic decay rates,
as well as checking a variety of hadron masses. So a complete programme of 
this kind encompasses the whole range of flavor physics. 
Figure~\ref{fig:ratio} shows a 2007 update of a range of quantities obtained 
from lattice QCD calculations with improved staggered sea quarks~\cite{highp}. 
The impressive agreement across the board provides strong confirmation that 
lattice QCD is accurately describing the real world when sea quarks 
are included. A companion plot of results in the quenched approximation
showing 10\% errors has now been dropped since there is no longer any 
point in attempting to produce quenched results for comparison. It is 
to be hoped that other formalisms will produce their version of this 
`ratio plot' soon. 

The way in which lattice QCD calculations are done was briefly described earlier. 
Here it is useful to describe further how the parameters of QCD are fixed. 
QCD describes an enormous range of physics with very few
parameters: a mass for each quark flavor and a coupling constant. For current 
lattice QCD calculations the $t$ quark is ignored and the $u$ and $d$ quarks are 
taken to have the same mass for numerical speed and convenience. This gives 
5 parameters to be fixed and we do this by fixing 5 hadron masses or mass differences 
to their experimental values. It is important to use gold-plated hadrons 
since hadrons that decay strongly or are close to decay thresholds will be 
sensitive to coupling to real or virtual decay channels that will distort the 
mass and, if there are systematic errors here, these will then be fed into the 
rest of the calculation. The hadron mass being used should be sensitive to the 
quark mass it is being used to fix but preferably not sensitive to other quark 
masses to avoid a complicated iterative tuning problem.  For the improved staggered 
results shown here the lattice spacing is fixed from the radial excitation energy 
in the $\Upsilon$ system, i.e. the difference in mass between the $\Upsilon^{\prime}$ and the 
$\Upsilon$, which turns out to be insensitive to all quark masses~\cite{alan}. The 
$u/d$ quark mass is then fixed from $M_{\pi}$, the $s$ quark mass from $M_K$~\cite{hisqf, toussaint}, the $c$
quark mass (using HISQ) from $M_{\eta_c}$~\cite{hisqf} and the $b$ quark mass (using NRQCD) 
from $M_{\Upsilon}$~\cite{alan}.   
Other gold-plated quantities can then be calculated with no free parameters 
and these are shown in Figure~\ref{fig:ratio}. Other choices to fix the 
parameters could be made, particularly for the lattice spacing itself. The ETMC 
collaboration advocate using $f_{\pi}$~\cite{urbach} 
and the RBC/UKQCD collaboration $m_{\Omega}$~\cite{boyle}. 
At intermediate points in the calculation it is convenient to use a nonphysical 
quantity to determine the relative lattice spacing very accurately between 
different ensembles. Most groups use a distance parameter from the heavy quark potential, 
either $r_0$ or $r_1$ corresponding to different values for the force between 
two infinitely massive quarks~\cite{sommer}. This can be determined with better than 0.5\% 
statistical accuracy but its physical value cannot be directly determined from 
experiment so at the end there must be conversion to physical units using, 
for example, $\Upsilon(2S-1S)$~\cite{alan}. 

Lattice QCD then provides a very natural and 
accurate way to determine the parameters of QCD, superior to any other method,
 and results from this have 
made their way into the particle data tables~\cite{pdg}. Further work on this is ongoing, but I will 
not report on it here.  
There will be a number of new results next year with improved accuracy for quark masses.  

\begin{figure}[h]
\rotatebox{270}{\includegraphics[height=8.0cm]{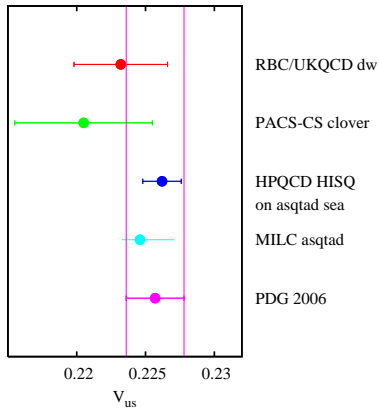}}
\caption{ Determinations of $V_{us}$ from lattice 
QCD results for $f_K/f_{\pi}$ combined with experimental 
results for the ratio of leptonic decay rates. 
The lattice QCD results use different formalisms~\cite{hisqf, bernard, juttner, kuramashi} but all 
with 2+1 flavors (i.e. $u/d$ and $s$) of sea quarks. 
}
\label{fig:vus}
\end{figure}

Here I will concentrate on lattice QCD calculations for CKM element determination 
starting with $V_{us}$.  The determination of $f_{\pi}$ was described above 
and there is an analogous calculation for the $K$ meson, yielding $f_K$. 
Again an `experimental' result for $f_K$ can be obtained from the experimental 
leptonic decay rate combined with a $V_{us}$ from elsewhere. Usually $V_{us}$ 
is taken from $K$ semileptonic decay which I will discuss shortly. This is 
the experimental result that is used in the ratio plot of Figure~\ref{fig:ratio}. 
(Note that the experimental value for $f_K$ used there has been updated from that 
quoted in the 2006 particle data tables~\cite{pdg} to be consistent with their quoted 
value of $V_{us}$).
The lattice results in that plot come from HISQ valence $u/d$ and $s$ quarks 
on the MILC improved staggered ensembles, with lattice errors of 1-2\%. 
The calculation of $f_K/f_{\pi}$ in lattice QCD can be done with a smaller 
error - 0.6\% for HISQ on the MILC ensembles - and this can be used, along 
with the ratio of the experimental leptonic decay rates~\cite{kloe, pdg}, to determine 
$V_{us}/V_{ud}$ and therefore $V_{us}$~\cite{marciano}.  In this way the HPQCD collaboration 
recently obtained $V_{us} = 0.2262(14)$~\cite{hisqf} 
and the MILC collaboration updated their previous $f_K/f_{\pi}$ analysis~\cite{milc3}
to give 0.2246(+25-13)~\cite{bernard}. Both are competitive with the result quoted from $Kl3$ decay 
in the particle data tables of 0.2257(21). Figure~\ref{fig:vus} shows a 
comparison of $V_{us}$ values obtained from recent lattice results for 
$f_K/f_{\pi}$ using 2+1 flavors of sea quarks. It is clear that this 
is an accurate way of determining $V_{us}$. The final result is still 
completely dominated by theoretical error, however. To improve it further 
will require a number of improvements to the lattice calculation, 
chiefly working on larger volumes to reduce the finite volume error in 
$f_{\pi}$ and reducing the uncertainty in the determination of the lattice 
spacing.

\begin{figure}[h]
\includegraphics[height=4.0cm]{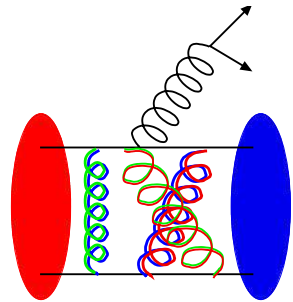}
\caption{ A 3-point function in lattice QCD involves 
two hadrons separated in time with the insertion of an 
operator in between to convert a quark from one flavor 
to another. Fitting this correlator as a function of 
the time position of the current, and of the time separation 
of the two hadrons, for various values of the spatial 
momenta of the two hadrons then allows the form factor 
as a function of squared 4-momentum transfer, $q^2$, to 
be determined.  
}
\label{fig:3pt}
\end{figure}

\begin{figure}[h]
\includegraphics[height=6.0cm]{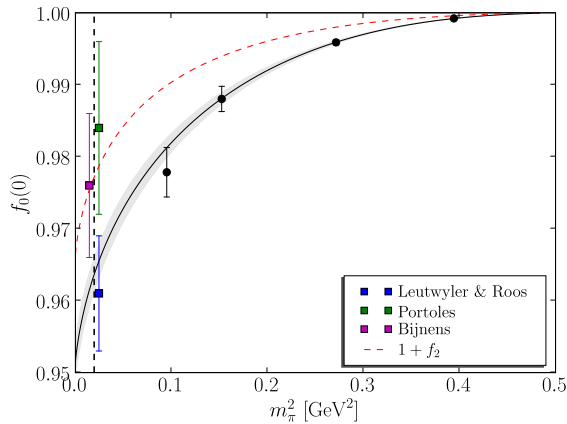}
\caption{ The scalar form factor, $f_0(0)$, for $K \rightarrow \pi$ decay 
obtained from lattice QCD calculations with 2+1 flavors of domain 
wall quarks~\cite{rbckl3}. The plot shows the form factor as a function of quark 
mass and fits used to extrapolate to the physical point. 
For comparison values are given from continuum estimates of 
various groups. 
}
\label{fig:dwkpi}
\end{figure}

\begin{figure}[h]
\includegraphics[height=7.0cm]{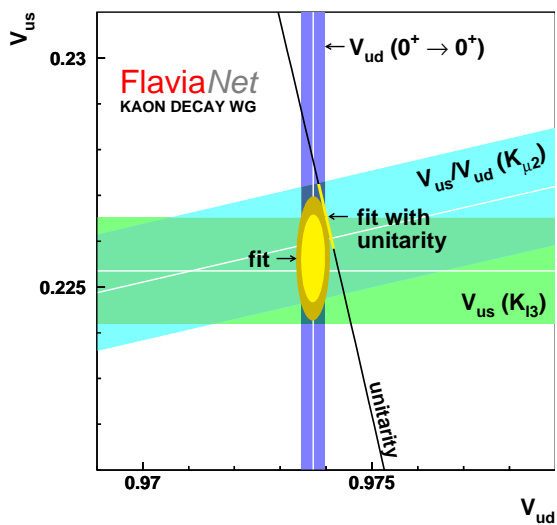}
\caption{ A combined fit by the Flavianet WG to test consistency 
of $V_{us}$ and $V_{us}/V_{ud}$ determinations using 
Kl3 and Kl2 decays along with lattice results. The constraint 
from first-row unitarity 
of the CKM matrix is also shown.
}
\label{fig:flavia}
\end{figure}

$V_{us}$ can also be determined from $K$ semileptonic decay to $\pi l \nu$. 
Because this is now a 3-body decay the quantity that must be 
determined in the lattice QCD calculation is a form factor that 
depends on $q^2$, the square of the 4-momentum transfer between the 
$K$ and the $\pi$. This involves calculating a so-called 3-point 
function, shown in Figure~\ref{fig:3pt}, where two different hadron 
operators are pulled apart in time and a current operator is inserted 
to convert a quark from one flavor to another. This is a much more 
complicated calculation than that of a 2-point function and in addition
must be done at a range of spatial momenta so that the $q^2$ dependence 
of the form factor can be extracted. The result required for $K \rightarrow \pi$ is the form 
factor at $q^2=0$ and this can be obtained either by extrapolation from 
results at the lattice spatial momenta available or 
by generating appropriate spatial momenta to give $q^2=0$ by creative 
use of the lattice boundary conditions. The RBC/UKQCD collaboration 
have new results this year on the $K \to \pi$ form factor using 
these techniques and the domain wall quark formalism~\cite{juttner}. The advantage of 
using $K$ semileptonic decay for $V_{us}$ is that the chiral extrapolation of the 
form factor is known to be relatively benign because of the Ademollo-Gatto 
theorem. This has been used in the past to estimate the difference 
of the form factor from 1.0 at $q^2=0$ (and this is what existing $V_{us}$ 
determinations are based on) but it is to be hoped that 
lattice QCD can give a more accurate result. The RBC/UKQCD collaboration 
find $f_+(0) = 0.9644(49)$ which yields $V_{us} = 0.2249(14)$~\cite{rbckl3}.  The error 
in the lattice result is estimated to be 0.6\% with results available 
currently at one value of the lattice spacing. Results from multiple values 
of the lattice spacing should allow this to be improved. Figure~\ref{fig:dwkpi} shows 
their results for $f_+(0)$ as a function of $u/d$ quark mass (given 
by $m_{\pi}^2$) and their chiral extrapolation compared to continuum 
estimates from different groups. 
Both the $Kl2$ and $Kl3$ results 
show that lattice QCD has an important role to play in the determination 
of $V_{us}$. This is summarised in the Flavianet plot~\cite{moulson}, Figure~\ref{fig:flavia}, 
that shows the $V_{us}$ from $Kl3$ (using lattice results from RBC/UKQCD~\cite{juttner}) and the ratio 
of $V_{us}/V_{ud}$ (using HPQCD lattice results~\cite{hisqf}) compared to first-row unitarity constraints.      

\begin{figure}[h]
\includegraphics[height=2.0cm]{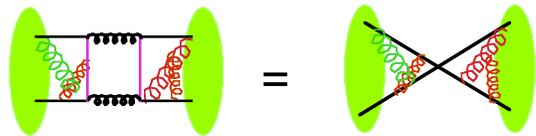}
\caption{ The `box diagram' reduces to a 4-quark operator 
in the effective weak Hamiltonian.  It is the matrix element 
of this latter operator that is calculated on the lattice.  
}
\label{fig:box}
\end{figure}

Neutral kaon mixing is an important process through which we can 
study CP violation in that system. In the Standard Model it proceeds
via the `box' diagram shown in Figure~\ref{fig:box} with $W$ boson 
exchange. On the lattice, we are working at relatively low energy scales 
compared to the $W$ boson mass and the version of the box diagram 
that is appropriate is that of a 4-quark operator from the effective 
weak Hamiltonian at low energy scales. The matrix 
element of this operator between neutral kaon states is parameterised 
by $f_K^2 B_K$ where $f_K$ is the decay constant and $B_K$, the amount 
by which the matrix element differs from $f_K^2$ is known as the bag factor. 
An evaluation of the 4-q operator matrix element in lattice QCD along 
with experimental results for the direct CP violation ratio of 
$2\pi$ decay rates of $K_L$ and $K_S$, $\epsilon$, allows 
a constraint on the CKM elements that enter the $4-q$ operator from 
$c$ and $t$ quarks appearing in the box diagram~\cite{buras}.  
There is no need in principle to separate the matrix element into its $f_K$ and 
$B_K$ components, but this is usually done and the two numbers quoted 
separately. 
$f_K$ on its own is a relatively simple calculation, as described above, 
and has been calculated accurately in lattice QCD by several groups. 
The 4-q operator is a harder calculation and is complicated to 
renormalise in formalisms that do not have the complete continuum 
chiral symmetry. This has made it a good calculation for the 
domain wall quark formalism. 

RBC/UKQCD have given a result for $B_K$ using domain wall quarks 
this year on their configurations including $u$, $d$, and $s$ 
sea quarks. They have calculations on both $16^3$ and $24^3$ lattices 
at a single lattice spacing of 0.11fm. 
They obtain 
$B_K^{\overline{MS}}(2 {\rm GeV})=0.524(10)(28)$ where the first
error is statistical and the second systematic~\cite{boyle}. 
However, their result for $f_K$ on the same lattices is currently 
5\% low. Encouraging preliminary results, but no final numbers, were also presented this year 
from a mixed-action approach using 
domain wall valence quarks on the MILC gluon configurations that 
include sea quarks with the improved staggered formalism at 
multiple values of the lattice spacing~\cite{ruth}.  Errors 
below 5\% are clearly possible on $B_K$ from lattice calculations 
in the near future. 

Charm physics provides a key test of lattice QCD calculations 
because, as described earlier, it is more sensitive to discretisation 
errors coming from the lattice QCD method than light quark physics. 
A new formalism called HISQ quarks~\cite{hisq} has shown this year that it is possible to 
treat charm quarks in essentially the same way as light quarks.
This makes for a very fast method (because it is based on the 
fast staggered formalism, but working at larger quark masses it 
is even faster) with the added advantage of being able to make 
use of light quark chiral symmetry to give conserved currents 
for calculating decay rates that do not need renormalisation (and 
their associated systematic errors). This then gives errors at the 
few percent level~\cite{hisqf}.

\begin{figure}[h]
\includegraphics[height=7.0cm]{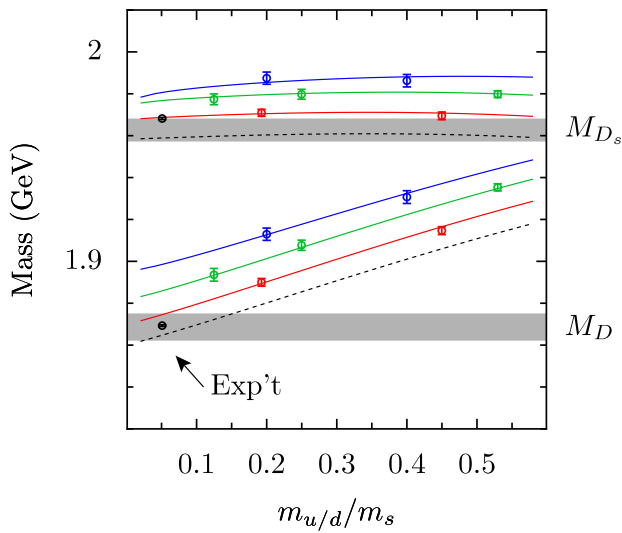}
\caption{ Masses of the $D^+$ and $D_s$ meson as a function 
of the $u/d$ mass in units of the $s$ mass at three values of the 
lattice spacing using the MILC ensembles. The lines gives the 
simultaneous chiral fits to all three lattice spacings and the 
dashed line the continuum extrapolation~\cite{hisqf}. The final error 
bars are given by the shaded bands, offset from the dashed lines 
by an estimate of electromagnetic, $m_u \ne m_d$ and other 
systematic corrections to the masses. The experimental results 
are marked at the physical $m_d/m_s$. 
}
\label{fig:Dspect}
\end{figure}

One significant test of the formalism, and one that has not been 
available from previous lattice calculations, is that of
the simultaneous determination of the spectrum of charmonium states and 
charm-light states with the same charm quark propagators. Of course, 
in QCD we know that there is only one charm quark with a given 
mass, but most approximations to QCD, such as potential models, 
find it impossible to handle both sets of states in the same 
approximation since their internal dynamics is very different. 
On the lattice it is also true that systematic errors are very 
different in the two systems, with charmonium states being more 
sensitive to discretisation errors, so that a very accurate discretisation 
is needed to be able to describe both successfully. With HISQ we have 
this and so
the value that the charm quark mass needs to take is fixed to 
get the $\eta_c$ mass correct (this being the lowest-lying charmonium 
state and one whose mass is most accurately calculated on the 
lattice). The $D_d$ and $D_s$ masses are then non-trivial {\it predictions} 
given a $u/d$ mass from $m_{\pi}$ and an $s$ quark mass from 
$m_K$. Very accurate results are obtained using valence HISQ quarks 
on the MILC ensembles, which agree with 
experiment with 6 MeV errors, see Figure~\ref{fig:Dspect}. 
This level of accuracy requires an understanding of corrections 
to the meson masses in the real world from QED effects and the 
fact that the $u$ and $d$ masses are not the same. To be working 
at the level of precision where these effects have to be considered 
is very exciting. 

\begin{figure}
\includegraphics[height=10.0cm]{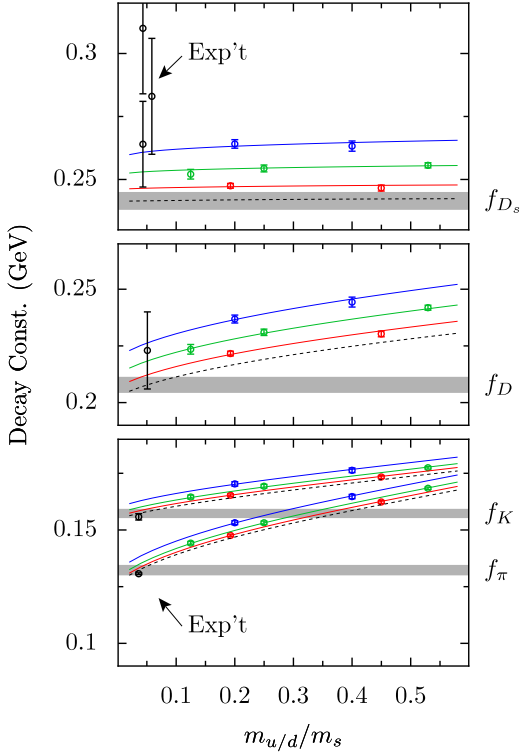}
\caption{ Results for the $D$, $D_s$, $K$ and $\pi$ 
decay constants on the very coarse, coarse and fine 
MILC ensembles using valence HISQ quarks. The chiral 
fits are performed simultaneously with those of the 
corresponding meson masses and the resulting continuum 
extrapolation curve is given by the dashed line. The
shaded band is the final result. Experimental results 
are shown on the left. For $K$ and $\pi$ these are 
from the PDG~\cite{pdg}. For $D$ and $D_s$ these 
are from CLEO-c~\cite{cleoc, cleocfd} (with $\mu$ and $\tau$ 
results shown separately) and BaBar~\cite{babar}.
}
\label{fig:fpifd}
\end{figure}

\begin{figure}[h]
\rotatebox{270}{\includegraphics[height=8.0cm]{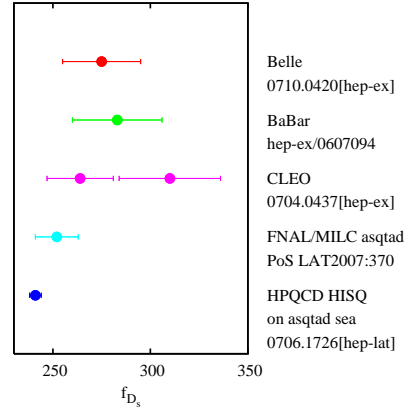}}
\caption{ A comparison of lattice results for $f_{D_s}$ from 
calculations including 2+1 flavors of sea quarks, compared 
to experimental determinations from the leptonic decay rate 
and using $V_{cs} = V_{ud}$. The two CLEO-c results from 
$\mu$ and $\tau$ channels are separated. 
}
\label{fig:fdscomp}
\end{figure}

The $D$ and $D_s$ decay constants can also be determined to an 
accuracy of 2\% using exactly the same method as for $f_K$ and 
$f_{\pi}$ described above. The results are shown in Figure~\ref{fig:fpifd}.
The leptonic decay rates of $D$ and $D_s$ mesons have now been measured 
by experiment and, given a value for the appropriate CKM element from 
elsewhere, can be converted into a value for the decay constant 
that can be compared to lattice results. Figure~\ref{fig:fdscomp} compares 
such experimental determinations of $f_{D_s}$ from BaBar, Belle and 
CLEO-c, using $V_{cs}=V_{ud}$, with lattice results. The Fermilab/MILC 
collaboration lattice result is also shown. They use the clover 
formalism for charm quarks on the MILC 
ensembles and produced first 
results for $f_D$ and $f_{D_s}$ ahead of experimental results and with errors 
of 7\%~\cite{fnalfold}. The value of $f_{D_s}$ shown in figure~\ref{fig:fdscomp} is an updated one from 
this year's lattice conference of 254(14) MeV~\cite{simonelat07}. The calculation of $f_D$ and 
$f_{D_s}$ in the clover formalism does require a renormalisation (and the 
error associated with this is included in the error estimate) and the 
mass of the charm quark in this case is fixed from the $D_s$ mass itself, 
so further independent checks of this calculation against other quantities 
need to be done. 
The two lattice calculations 
agree well although they use very different charm quark formalisms, albeit on 
the same gluon ensembles. Further work is underway from other groups using 
other formalisms and ensembles. 

On this quantity the lattice results are ahead of experiment, although
experimental errors are expected to improve by a factor of two over the 
coming year. The experimental central values will have to come down as that 
happens if there is to be agreement. One outstanding issue is that of 
electromagnetic effects in the experimental result since the decay 
constant is defined in pure QCD~\cite{marcsir}. More work needs to be done 
in the $D$ and $D_s$ cases to ensure that is not an issue at the level of 
precision we are aiming for in this comparison between theory and experiment.

\begin{figure}[h]
\includegraphics[height=6.0cm]{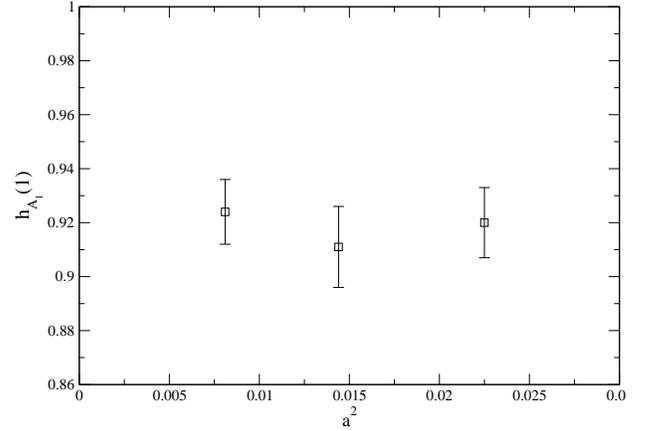}
\caption{The form factor at zero recoil for $B \rightarrow D^* l \nu$ 
calculated by the Fermilab/MILC collaboration on the MILC configurations
at three values of the lattice spacing (in ${\rm fm}^2$. 
Good consistency between them is observed. The final 
result is given as 0.924(12)(19)~\cite{laiho}. 
}
\label{fig:fnalf1}
\end{figure}

Neutral $B$ mesons can mix in an analogous way to $K$ mesons. 
In this case it is simpler, being dominated by a single 
box diagram with $t$ quarks in, and reducing to a 4-quark operator with 
coefficient $V_{td/s}V_{tb}^*$ (see Figure~\ref{fig:box}). The matrix element of this operator 
is parameterised by $f_B^2B_B$ where $f_B$ is the decay constant. 
In the past there has been a lot more work on the decay constant than 
on the 4-quark operator. This year, however, two groups, the HPQCD 
collaboration and the Fermilab/MILC collaboration, presented new 
results for the ratio of the 4-quark operator matrix element for 
the $B_s$ to that of the $B_d$~\cite{gamiz}. The results are at a preliminary 
stage currently but will be improved significantly over the 
coming year. The ratio $\xi = f_{B_s}\sqrt{B_{B_s}}/f_B\sqrt{B_B}$ is 
more accurate than the individual numbers because we cannot escape 
a renormalisation (and its associated error) of the matrix elements in this case, but the 
renormalisation is the same for $B$ and $B_s$. Previously the 
HPQCD collaboration gave a result for $f_{B_s}/f_B$ of 1.20(3)~\cite{ourblept}. This 
year the Fermilab/MILC collaboration give their value for 
this ratio as 1.22(3)~\cite{simonelat07}.  Our 
prejudice is that $\xi$ should be very similar to $f_{B_s}/f_B$, and the 
aim is certainly to achieve the same level of error, or better. 
HPQCD have given a result for $f_{B_s}\sqrt{\hat{B}_{B_s}}$ of 
0.281(21) GeV~\cite{bsmix}, where a lot of this error comes from the renormalisation. 

Exclusive $B$ semileptonic decay is an important route to determining 
the CKM elements $V_{cb}$ and $V_{ub}$. This year the Fermilab/MILC 
collaboration have determined the appropriate form factor
for the process $B \rightarrow D^* l \nu$, including the effect 
of sea quarks fully for the first time~\cite{laiho}. The rate of this decay at zero 
recoil is proprotional to the square of $V_{cb}\times h_{A_1}(1)$, where $h_{A_1}(1)$ 
is the form factor for $B$ and $D^*$ relatively at rest. Figure~\ref{fig:fnalf1}
shows the Fermilab/MILC results for $h_{A_1}(1)$ at three 
different values of the lattice spacing
using the MILC gluon field configurations including the effect of sea quarks. 
From their results they determine $h_{A_1}(1) = 0.942(12)(19)$ where the first error 
is statistical and the second systematic. Using the HFAG experimental average~\cite{hfag}, 
this leads to a value for $V_{cb}$ of $38.7(0.7)(0.9) \times 10^{-3}$ where the 
first error is from experiment and the second from the lattice. 

$V_{ub}$ can be extracted from the exclusive $B \rightarrow \pi l \nu$ 
process. Here lattice results are still not very accurate because the 
important kinematic region is one in which the $\pi$ mesons are moving 
rather fast and this gives a very noisy signal on the lattice. Work 
is underway to ameliorate this~\cite{wonglat07}. A recent theoretical determination of 
$V_{ub}$ using combined lattice results from HPQCD and Fermilab/MILC along 
with light-cone sum rules gives $V_{ub} = 3.47(29) \times 10^{-3}$~\cite{flynn}. 
The issue of compatibility of this result with that from the 
inclusive $b \rightarrow u$ decay is becoming an important one~\cite{neubert}. 

\section{Conclusions}

Lattice calculations including the full effect of $u$, $d$ and 
$s$ sea quarks are in excellent shape. Calculations using improved 
staggered quarks continue to get better and new results are now 
appearing from other quark formalisms. There have been significant 
new results this year in strange and charm physics. I have 
concentrated on results relevant to CKM physics and have not mentioned 
many other areas of progress in lattice calculations. The reader 
is referred for those to the proceedings of this year's lattice conference~\cite{lat07}. 

\begin{figure}[h]
\includegraphics[height=8.0cm]{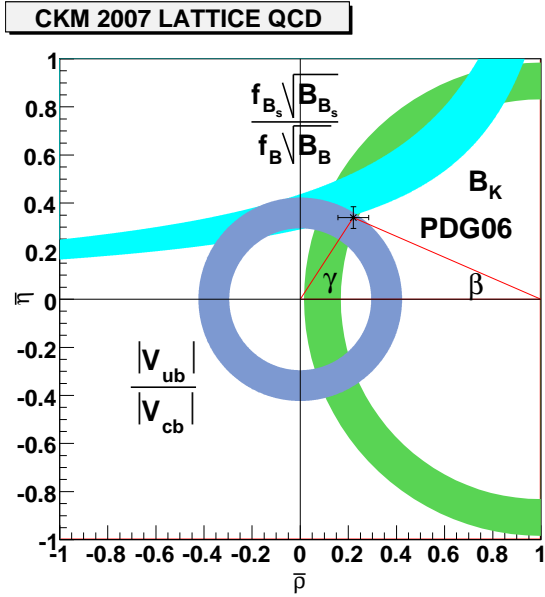}
\caption{Constraints on the unitarity triangle from the 
CKM matrix using current lattice results that include 
the effect of $u$, $d$ and $s$ sea quarks. The 
lattice errors dominate these constraints and they 
can be halved over the next two years. The cross gives 
the current constraint on the vertex quoted in the 
PDG~\cite{pdg}. 
}
\label{fig:ckmtri}
\end{figure}

In Figure~\ref{fig:ckmtri} I have collected recent lattice results into a 
set of constraints on the upper vertex of the standard 
unitarity triangle plot.  The lattice inputs needed are 
$B_K$, $f_K/f_{\pi}$, $f_+(K \rightarrow \pi l \nu)$, 
$F(B \rightarrow D^* l \nu)$, $f_+(B \rightarrow \pi l \nu)$ 
 and $f_{B_s}\sqrt{B_{B_s}}/f_B\sqrt{B_B}$. It is important 
to use lattice results from the calculations including
the full effect of $u$, $d$ and $s$ sea quarks. Old results 
in the quenched approximation are not reliable enough for 
this. They have in any case now been superseded and should not be used. 

In the next two years the lattice errors on these CKM constraints 
should halve. The robustness of our error estimates will be further 
tested against experiment using other gold-plated results. 
$\Gamma_{e^+e^-}$ for $\psi$ and $\Upsilon$ are good tests for $c$ and $b$ 
physics, for example. We have no 
free parameters when we do this and so it is a very stringent test. The era 
of precision lattice QCD calculations has really arrived.  

{\bf Acknowledgements}. I am grateful to the conference organisers for the 
opportunity to give this talk and to the following people for assistance 
in preparing it: Claude Bernard, Peter Boyle, Eduardo Follana, Elvira Gamiz, Andreas J\"{u}ttner, 
Andreas Kronfeld, Jack Laiho, Peter Lepage, Paul Mackenzie, Craig McNeile, Matthew Moulson, Gerrit 
Schierholz, Enno Scholz, Junko Shigemitsu, Jim Simone, Doug Toussaint, Carsten Urbach, Ruth van de Water 
and Kit Wong.

\end{document}